%% file: wcl-pipecsi.tex
\documentclass[journal]{IEEEtran}

\input{macros}

\usepackage{balance}
\usepackage{float}
\begin{document}
	\title{Model Splitting Enhanced Communication-Efficient Federated Learning for CSI Feedback}
	\author{Yanjie Dong, Haijun Zhang,~\IEEEmembership{Fellow, IEEE}, Gaojie Chen,~\IEEEmembership{Senior Member, IEEE}, Xiaoyi Fan, 
		\mbox{Victor C. M. Leung,~\IEEEmembership{Life Fellow, IEEE}}, and Xiping Hu
		\thanks{This work was supported by the National Natural Science Foundation of China under Grant 62102266 and the Pearl River Talent Recruitment Program of Guangdong Province under Grant 2019ZT08X603,
			Public Technology Platform of Shenzhen City (GGFW2018021118145859), 
			Shenzhen Science and Technology Innovation Commission (R2020A045, KCXFZ20201221173411032).}
		\thanks{Y. Dong, V. C. M. Leung, and X. Hu are with the Artificial Intelligence Research Institute and Guangdong-Hong Kong-Macao Joint Laboratory for Emotional Intelligence and Pervasive Computing, Shenzhen MSU-BIT University, Shenzhen 518172, China.}
		\thanks{H. Zhang is with the Beijing Engineering and Technology Research Center for Convergence Networks and Ubiquitous Services, University of Science and Technology Beijing, Beijing 100083, China.}
		\thanks{G. Chen is with the School of Flexible Electronics \& State Key Laboratory of Optoelectronic Materials and Technologies, Sun Yat-Sen University, Shenzhen 518107, China.}
		\thanks{X. Fan is with Department of Computer Science and Engineering, The Hong Kong University of Science and Technology, Hong Kong, and Jiangxing Intelligence Inc., Shenzhen, China.}
	}
	
	\maketitle
	
	\begin{abstract}
		Recent advancements have introduced federated machine learning-based channel state information (CSI) compression before the user equipments (UEs) upload the downlink CSI to the base transceiver station (BTS).
		However, most existing algorithms impose a high communication overhead due to frequent parameter exchanges between UEs and BTS. 
		In this work, we propose a model splitting approach with a shared model at the BTS and multiple local models at the UEs to reduce communication overhead. 
		Moreover, we implant a pipeline module at the BTS to reduce training time. 
		By limiting exchanges of boundary parameters during forward and backward passes, our algorithm can significantly reduce the exchanged parameters over the benchmarks during federated CSI feedback training.
	\end{abstract}
	
	\begin{IEEEkeywords}
		Communication efficiency, CSI codebook learning, federated training, pipeline parallelism.
	\end{IEEEkeywords}
	
	\section{Introduction}
	The massive multi-input and multi-output (mMIMO) technology has been standardized in the fifth generation (5G) mobile communications \cite{Chen2023, Xu2023}. 
	Equipped with a large-scale antenna array, an mMIMO system can effectively suppress multi-user interference and significantly enhance system capacity by harnessing spatial diversity and multiplexing gains.
	However, during the downlink period,  the merits of the mMIMO system depend on accurate acquisition of channel state information (CSI) at the base transceiver station (BTS) \cite{Liu2024}. 
	While user equipments (UEs) can accurately estimate the downlink CSI \cite{Verenzuela2020, Gao2024}, they need to upload the CSI estimates to the BTS via the feedback channels in the frequency division duplex (FDD) mode.
	When the scale of the antenna array continuously increases, downlink CSI feedback can incur a burdensome overhead that may compromise the benefits of mMIMO technology \cite{Guo2022a}. 
	Therefore, the downlink CSI needs to be compressed before uploading to the BTS. 
	
	Three categories of compression algorithms have been proposed to reduce the CSI feedback overhead, i.e., codebook based algorithms \cite{Dreifuerst2023, 3gppcodebook},  compressive-sensing based algorithms \cite{Liang2020}, and deep-learning based algorithms \cite{Wen2018, Ji2021}. 
	More specifically, with a shared codebook, the UEs only need to upload the indices of codewords to the BTS to recover the downlink CSI. However, the length of codewords linearly increases with the number of antennas. 
	On the other hand, compressive-sensing based algorithms depend on the sparsity of downlink CSI, which  may not hold when advanced multiplexing schemes are used \cite{Chi2024}.
	Besides, the iterative procedures of compressive-sensing based algorithms also hinder the timeliness of acquiring downlink CSI at the BTS \cite{Wen2018, Ji2021} and the downstream tasks \cite{Dong2022, Li2024, Dong2024}. 
	Recent advancements in deep learning have  promoted the use of machine learning models for compressing downlink CSI \cite{Wen2018, Ji2021}.
	For example, in CsiNet \cite{Wen2018}, CLNet \cite{Ji2021}, and Transformers \cite{Wang2023, Huang2025}, deep neural networks are trained before deployed at the UEs and the BTS. 
	Moreover, the correlations of different data domains \cite{Zhang2023a} and radio resource management based on limited CSI feedback \cite{Sohrabi2021, Gao2022} are exploited in the context of deep learning based CSI feedback.
	
	While deep-learning based approaches for compressing CSI feedback have received significant attention from both industry and academia, the previous deep neural networks and their associated training methodologies have been primarily developed for the CSI feedback of a single UE. 
	To address this limitation, deep neural networks have been proposed to facilitate downlink CSI feedback for multiple UEs \cite{Mashhadi2021, Guo2024}. Furthermore, distributed training algorithms, which are the focus of our work, have been explored using frameworks such as federated learning \cite{Cui2024} and gossip learning \cite{Guo2022}.
	However, the previous distributed training algorithms require each UE to either upload the full model parameters to the BTS \cite{Cui2024} or to exchange full parameters with the neighbour UEs \cite{Guo2022}. 
	Full-parameter exchanges can introduce a communication bottleneck when the number of UEs becomes large.

	Different from the previous distributed training algorithms, we propose a new communication-efficient model splitting algorithm that is named as \textbf{CSILocal} to train the CSI feedback deep neural networks (DNNs). 
	Our key contributions are summarized as follows. 
	\begin{itemize}
		\item To reduce the communication expenditure, the proposed CSILocal algorithm divides the CSI model into three modules, i.e., encoder, decoder tail, decoder head. 
		Different from traditional distributed training algorithms that need to exchange the full/partial model parameters, our proposed CSILocal algorithm only needs to exchange the local smashed data that is significantly less than the size of the CSI feedback DNN.
		In addition, the original CSI data are kept at each UE to preserve the privacy of UEs.
		\item We further split the decoder tail into several modules for pipeline parallelism such that the training duration can be further reduced. 
		\item 	Numerical experiments are conducted to demonstrate that our proposed CSILocal algorithm can significantly reduce the communication costs between the UEs and BTS. 
		Moreover, numerical results also verify that our proposed pipeline parallelism can reduce the training time duration.
	\end{itemize}

	\emph{Notations.} 
	Vectors and matrices are respectively denoted by lowercase- and uppercase-boldface letters. 
	$\mathbb{C}$ and $\mathbb{R}$ respectively represent the sets of complex and real values.  
	The operator $\bm X^{\dag}$ denotes the conjugate transpose of matrix $\bm X$.

	\section{System description}
	\subsection{Communication Signaling}
	We consider the downlink transmission of an mMIMO system with a single BTS and $N$ UEs.
	The BTS has $N_t$ transmit antenna, and each UE has one single antenna.
	The BTS communicates with the UEs over $N_c$ subcarriers. 
	Let $\bm H_n = [\bm h_{n,1}, \bm h_{n,2}, \ldots, \bm h_{n,N_c}] \in \cc^{N_t \times N_c}$ and $\bm W_n = [\bm w_{n,1}, \bm w_{n,2}, \ldots, \bm w_{n,N_c}] \in \cc^{N_t \times N_c}$ denote the downlink CSI matrix and precoding matrix of UE $n$, respectively. 
	When the input data streams of each UE $n$ are segmented into $N_c$ parallel streams as $\bm x_n \in \cc^{N_c \times 1}$, the received signal at UE $n$ is
	\begin{equation}\label{eqa:signal}
		\bm y_n = \bm H_n^{\dag} \bm W_n \bm x_n + \bm H_n^{\dag} \sum_{i \neq n}^{N}   \bm W_i \bm x_i + \bm z_n
	\end{equation}
	where $\bm z_n \in \mathbb{C}^{N_c\times 1}$ denotes the additive white Gaussian  noise with each entry being independent and identically distributed (i.i.d.) complex Gaussian with mean zero and variance $\sigma^2$.
	
	\begin{remark}
		The signal model in \eqref{eqa:signal} demonstrates that the design of the precoding matrix $\bm{W}_n$ critically depends on the accurate estimation of the downlink CSI matrices $[\bm{H}_n]_{n=1}^N$  as shown in \cite{Dong2022, Dong2024, Gao2022}.
		When the mMIMO system operates in FDD mode, each UE $n$ can estimate the downlink CSI matrix $\bm H_n$ via the pilot-based channel estimation.
		The estimated CSI matrix is uploaded to the BTS via the feedback channels.
		However, the communication overhead associated with CSI feedback becomes significantly burdensome as the number of antennas and subcarriers increases.
		Therefore, we are motivated to train a DNN-based encoder to compress each CSI matrix $\bm H_n$ into a codeword per each UE $n$ and a DNN-based decoder  to recover the compressed CSI matrices $[\bm H_n]_{n=1}^N$ at the BTS. 
	\end{remark}

	\subsection{Model Splitting}
	Each UE can hold $M$ private downlink CSI matrices and would not want to share them with the other entities.
	To preserve data privacy, the CSI feedback DNN needs to have a personal part at each UE and a shared part at the BTS to improve the training efficiency. We consider to split the model into three parts: the feature encoder and decoder head at each UE, and the decoder tail at the BTS.
	The UEs communicate with the BTS via digital transmission under the channel capacity. 
		In this case, the communication link between each UE and the BTS experiences a negligible error rate \cite{Yao2024}. 
		We assume that all UEs have identical processing capabilities; therefore, the impact of stragglers can be ignored under digital transmission.

	\begin{figure}[tb]
		\centering
		\includegraphics[width=\linewidth]{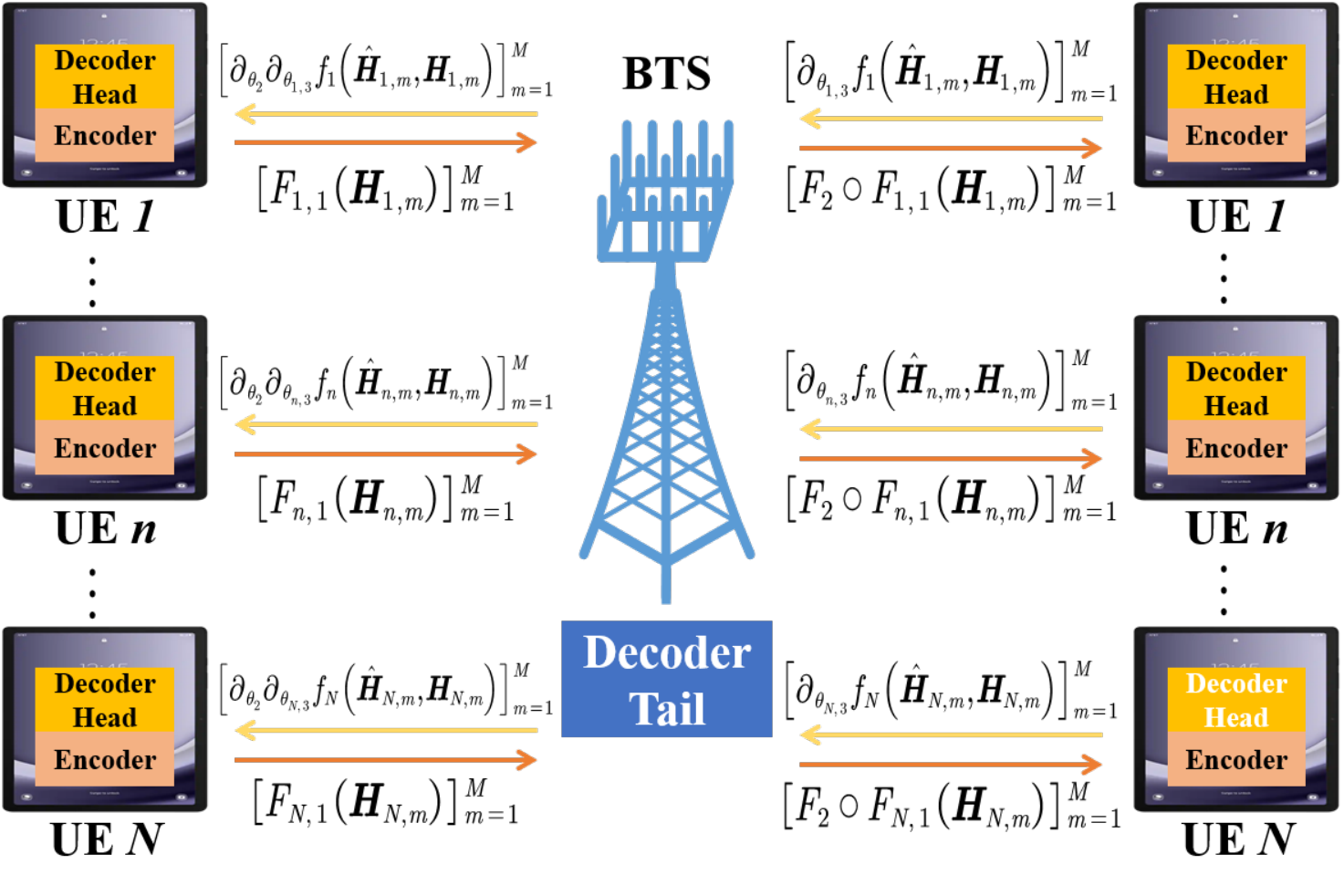}
		\caption{The model is split into three parts with the encoder and decoder head at each UE and the decoder tail at the BTS.}
		\label{fig:model-split}
	\end{figure}

	As shown in Fig. \ref{fig:model-split}, we consider the federated CSI feedback learning where a BTS and $N$ UEs coordinately work to minimize the normalized mean squared error as
	\begin{equation}\label{eq:mt}
		\min_{\Theta} \sum_{n=1}^N \sum_{m=1}^M  f_n(\hat{\bm H}_{n,m}, \bm H_{n,m})
	\end{equation}
	where $f_n(\hat{\bm H}_{n,m}, \bm H_{n,m}) = f(\bm\theta_{n,1}, \bm\theta_2, \bm\theta_{n,3}; \bm H_{n,k}) := \nicefrac{\norm{ \hat{\bm H}_{n,m} - \bm H_{n,m}}^2}{\norm{\bm H_{n,m}}^2}$ with $\hat{\bm H}_{n,m} = F_{n,3}  \circ F_{2} \circ F_{n,1} (\bm H_{n,m})$ denoting the $m$th reconstructed CSI matrix of UE $n$. 
	The encoder $F_{n,1}: \mathbb{R}^{2 \times N_t \times N_c} \rightarrow \mathbb{R}^{c_1}$, the decoder tail $F_{2}: \mathbb{R}^{c_1} \rightarrow \mathbb{R}^{c_2}$, and the decoder head $F_{n,3}: \mathbb{R}^{c_2} \rightarrow \mathbb{R}^{2 \times N_t \times N_c}$ are respectively parameterized by $\bm\theta_{n,1} \in \mathbb{R}^{d_1}$, $\bm\theta_{2}  \in \mathbb{R}^{d_2}$, and $\bm\theta_{n,3}  \in \mathbb{R}^{d_3}$ with ${\Theta} = \{ [\bm\theta_{n,1}, \bm\theta_{n,3}]_{n=1}^N, \bm\theta_2 \}$ and  $d = d_1 + d_2 + d_3$.
	{\color{black}
	\begin{remark}
	As shown in Fig. \ref{fig:model-split}, the BTS needs to exchange local smashed data with the $N$ UEs. 
	More specifically, the boundary activation tensors $[F_{n,1}(\bm H_{n,m})]_{m=1}^{M}$ and $[F_2 \circ F_{n,1}(\bm H_{n,m})]_{m=1}^{M}$ are exchanged between the BTS and UE $n$ during the forward pass. 
	During the backward pass, the upper-layer gradients $[\partial_{\theta_{n,3}} f_n({\hat{\bm H}}_{n,m}, {\bm H}_{n,m})]_{m=1}^{M}$ and $[\partial_{\theta_{2}} \partial_{\theta_{n,3}} f_n({\hat{\bm H}}_{n,m}, {\bm H}_{n,m})]_{m=1}^{M}$ are exchanged between the BTS and UE $n$.
	\end{remark}}
	
	\begin{remark}
		To reduce the communication expenditure between the BTS and the UEs, the dimensions at the sub-model boundaries (i.e., $c_1$ and $c_2$) should satisfy $c_1 \ll \min\{d_1, d_2\}$ and $c_2 \ll \min\{d_2, d_3\}$. 
	\end{remark}

	\section{CSILocal Algorithm}
	\begin{figure*}[!h]
		\centering
		\begin{minipage}{0.38\textwidth}
			\centering
			\includegraphics[height=3.5 cm]{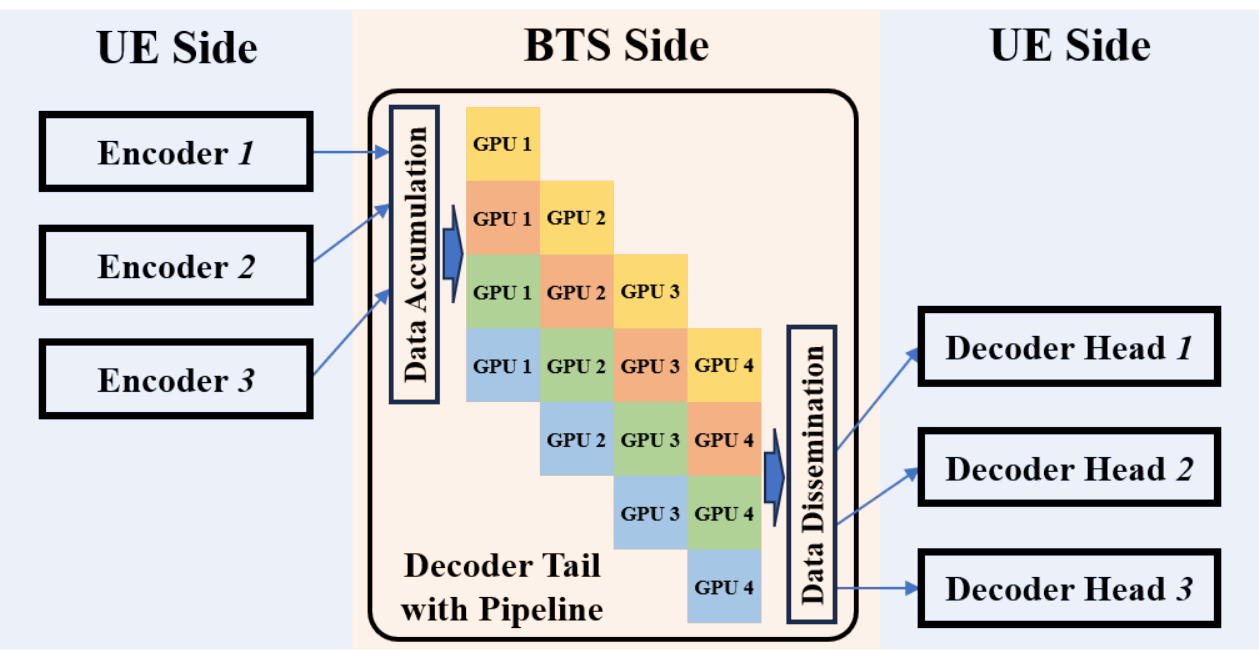}
			\caption{An illustration of the pipeline parallelism for the federated CSI feedback learning with three UEs.}
			\label{fig:csi-model-parallel}
		\end{minipage}%
		\hspace{1.5 pt}
		\begin{minipage}{0.6\textwidth}
			\centering
			\includegraphics[height=3.5 cm]{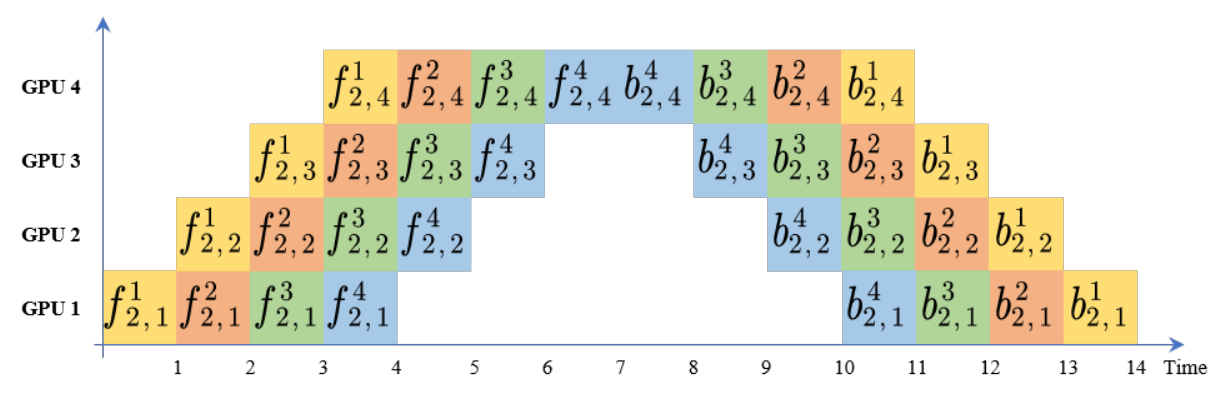}
			\caption{An illustration of pipeline parallelism on the forward and backward passes when the BTS has four GPUs.}
			\label{fig:fbpass}
		\end{minipage}
	\end{figure*}
	
	For the model training problem \eqref{eq:mt}, each UE $n$ can calculate the local loss values during the forward passes and the gradients for the three parts of models during the backward passes. 
	For each UE $n$, the gradients with respect to the encoder $\bm\theta_{n,1}$, decoder tail $\bm\theta_2$, and decoder head $\bm\theta_{n,3}$ are respectively denoted by 
	\begin{align}\label{eq:03}
		\nabla_{n,1} &:= \nabla_{\bm\theta_{n,1}} f_n(\hat{\bm H}_{n,m}, \bm H_{n,m}) \\
		\nabla_{2} &:= \sum_{n=1}^N \nabla_{\bm\theta_{2}} f_n(\hat{\bm H}_{n,m}, \bm H_{n,m}) 
	\end{align}
	and
	\begin{equation}\label{eq:04}
		\nabla_{n,3} := \nabla_{\bm\theta_{n,3}} f_n(\hat{\bm H}_{n,m}, \bm H_{n,m})
	\end{equation}
	where $m$ is uniformly drawn from the set $\{1, \ldots, M\}$.
	
	By leveraging the Adam optimizer \cite{Defossez2022}, each part of the model can be updated  per iteration $k$ as
	\begin{align}
		\bm\theta_{n,1}^{k+1} &=  \bm\theta_{n,1}^{k} - \eta \adam{\nabla_{n,1}^k, \bm\theta_{n,1}^{k}}\label{eq:05a} \\
		\bm\theta_{2}^{k+1} &=  \bm\theta_{2}^k - \eta  \adam{\nabla_{2}^k, \bm\theta_{2}^k}\label{eq:05b}
	\end{align}
	and
	\begin{equation}\label{eq:06}
		\bm\theta_{n,3}^{k+1} = \bm\theta_{n,3}^k  - \eta \adam{\nabla_{n,3}^k, \bm\theta_{n,3}^k} 
	\end{equation}
	where $\eta > 0$ is the learning rate.


	{\color{black}
	Based on the pipeline parallelism in Fig. \ref{fig:csi-model-parallel}, we summarize the procedures of  proposed \textbf{CSILocal} in \textbf{Algorithm \ref{alg:01}}.
	
	\begin{algorithm}[htb] 
		\centering\small \color{black} 
		\caption{CSILocal Algorithm}\label{alg:01}
		\begin{algorithmic}[1]
			\State Each UE $n$ initializes an Adam optimizer with learning rate $\eta$ and momentum factors $(\beta_1, \beta_2)$
			\For{$k = 1, 2, \ldots, K$}
			\State Each UE $n$ feeds the CSI matrix $\bm H_{n,m}$ into the encoder $n$ and uploads the smashed data of encoder $n$ to the BTS
			\State The BTS accumulates all local smashed data from the $N$ UEs via the data accumulation module 
			\State The BTS splits the accumulate data into $4$ micro batches as shown in Fig. \ref{fig:csi-model-parallel}
			\State The BTS processes  the $4$ micro-batches of smashed data via the decoder tail with pipeline as shown in Fig. \ref{fig:csi-model-parallel}
			\State The BTS disseminates the output of the decoder tail with pipeline to all UEs
			\State Each UE $n$ calcualtes the loss value based on the output of decoder head and the label of CSI matrix $\bm H_{n,m}$ 
			\State Each UE $n$ performs backward pass to calculate the gradients with respect to encoder, decoder tail and decoder head as \eqref{eq:03}--\eqref{eq:04}
			\State Each UE $n$ updates the parameters of encoder and decoder head via \eqref{eq:05a} and \eqref{eq:06}
			\State The BTS updates the parameters of decoder tail via \eqref{eq:05b}
			\EndFor
		\end{algorithmic}
	\end{algorithm}}

	{\color{black}
		Note that the volume of accumulated local smashed data at the BTS scales linearly with the number of UEs.
		When the number of UEs is getting large, the pipeline parallelism is required to handle the large volume of local smashed data at the BTS as shown in Fig. \ref{fig:fbpass}. 
		When the BTS has four GPUs, the decoder tail $F_2$ is divided into four sub-models to four GPUs, i.e., $F_2 = f_{2,4}\circ f_{2, 3} \circ f_{2, 2} \circ f_{2, 1}$. 
		At the data accumulation module of Fig. \ref{fig:csi-model-parallel}, the three mini-batch of smashed data are divided into four smaller micro batches.
		As shown in Fig. \ref{fig:fbpass}, GPU 2 can calculate the activation tensors of the first micro-batch of smashed data while the GPU 1 works on the calculation of forward activation tensors of the second micro-batch. As time goes by, the four GPUs can be used simultaneously to process the four micro-batches of smashed data. These procedures can be repeated during the backward passes. 
		At the data dissemination module of Fig. \ref{fig:csi-model-parallel}, the outputs of decoder tail are then divided into three mini-batch of smashed data and delivered to the corresponding three decoder heads.  
		The corresponding backward function can be computed via the automatic symbolic differentiation \cite{Huang2019}.
	}The introduced pipeline parallelism can reduce the consumed time for calculating the gradient $\nabla_2^k$ by improving GPU utilization efficiency on the BTS per each iteration $k$.
	In this way, the overall wallclock time for training can be reduced.

	\section{Numerical Results}	
	\subsection{Neural Network and Hyper-Parameters}
\begin{figure}[!htb]
		\centering
		\includegraphics[height= 4.1 cm]{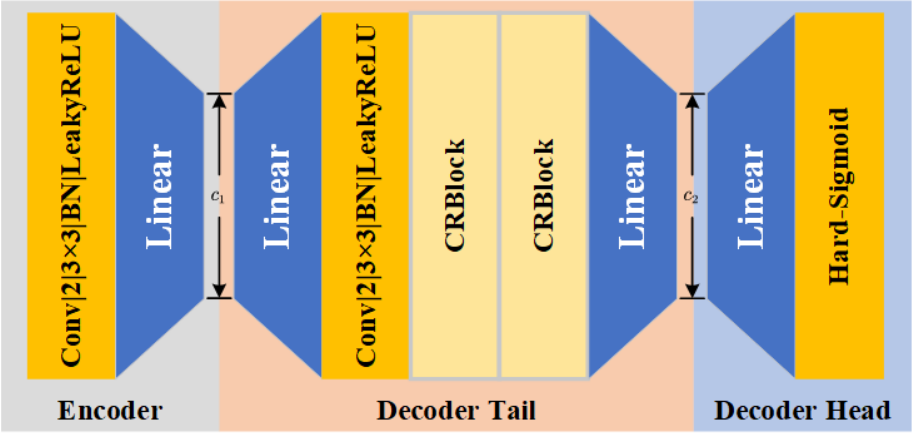}
		\caption{An illustration of used deep neural network that have encoder, decoder tail, and decoder head.
			Each UE is equipped with encoder and decoder head, and the BTS is equipped with the decoder tail. In order to reduce the communication overhead, the encoding dimension $c_1 = c_2$ are set smaller than the dimension of each data sample.}
		\label{fig:cnn-structure}
	\end{figure}

	\textbf{Description of neural network.} To facilitate the communication efficiency, we reconstruct the downlink CSI matrices via an autoencoder-based DNN that consists of encoder, decoder tail, and decoder head. 
	The proper padding is used to ensure the output of each convolutional layer match the size of the CSI matrix. As shown in Fig. \ref{fig:cnn-structure}, the detailed information of the three parts is as follows.
	
	\begin{itemize}
		\item \textbf{Encoder:} We reuse the encoder structure of CsiNet where the layer sequence is a convolutional layer with $2$ kernels and kernel size as $3\times 3$, a batch-normalization layer, a leaky rectified linear (LeakyReLU) activation with negative slope as $0.3$, and a fully connected layer. 
		\item \textbf{Decoder Tail:} To enhance the performance of reconstruction, the decoder tail consists of two fully connected layers, a convolutional layer with $2$ kernels and kernel size as $3\times 3$, a LeakyRelu activation with negative slope as $0.3$, and two CRBlocks \cite{Lu2020} with kernel sizes as $1\times 3$ and $1\times 5$.
		\item \textbf{Decoder Head:} The decoder head uses the fully connected layer to recover the structure of the CSI matrices and leverage the hard-Sigmoid activation to reconstruct the CSI matrices.
	\end{itemize}

	\textbf{Hyper-parameter setup.}~{\color{black}We consider six benchmarks: FedAvg, FedAvgPer, FedProx, FedProxPer, FedGrad, and FedGradPer \cite{Pillutla2022, Xu2023}. 
		More specifically, the FedAvg, FedGrad, and FedProx algorithms require the exchange of full model parameters (or gradients) between the UEs and the BTS in each iteration. 
		In contrast, the FedAvgPer, FedGradPer, and FedProxPer algorithms only require the exchange of the model parameters (or gradients) of the decoder heads and tails between the UEs and the BTS per each iteration.}
	For fair comparison, we use the same dataset as the CsiNet \cite{Wen2018}. 
	The generated CSI matrices are converted to angular-delay domain via the two-dimensional discrete fourier transform. 
	We combine the original $100,000$ training and $30,000$ validation samples are the training set, and use the remaining $20,000$ samples as the testing set. 
	Each data sample has size $32 \times 32$. 
	Unless otherwise specified, the mMIMO system is configured with $10$ UEs.
	We set the training iterations as $20,000$, the learning rate as $8\times10^{-5}$, and the momentum-factor tuple for the Adam optimizer as $(0.9, 0.95)$.
	{\color{black}
	Since we have limited number of GPUs, we set the number of micro batches as $2$. 
	We use the normlized mean-squared error loss. 
	Unless otherwise specified, we set the mini-batch size as $800$, and the dimensions of sub-model boundaries $c_1$ and $c_2$ as $256$.}
	
	{\color{black}
	\textbf{Non-IID setup.}~In order to demonstrate the impact of data heterogeneity in CSI data, we construct a synthetic dataset consisting of $130,000$ training samples and $20,000$ testing samples. 
	More specifically, the ratio of indoor to outdoor CSI data is set to $1:1$ in both the training and testing sets. 
	In the mMIMO system, each UE is assigned $13,000$ data samples, and the indoor-to-outdoor CSI ratios for the $10$ UEs are set as follows: $9.5:0.5$, $8.5:1.5$, $7.5:2.5$, $6.5:3.5$, $5.5:4.5$, $4.5:5.5$, $3.5:6.5$, $2.5:7.5$, $1.5:8.5$, and $0.5:9.5$.
	}
	
	\subsection{Pipeline vs. No Pipeline}
	\begin{table}[ht]
		\centering
		\caption{Duration per iteration  for training CSILocal model with and without pipeline parallelism at the server (measured in seconds).}
		\begin{tabular}{|l|c|c|c|c|}
			\toprule
			\hline
			Environments & \multicolumn{4}{c|}{Indoor} \\
			\hline
			Encoding Dimension & \multicolumn{2}{c|}{256} & \multicolumn{2}{c|}{512} \\
			\hline
			Batch Size per UE & 400   & 800   & 400   & 800 \\
			\hline
			CsiLocal with Pipeline & 0.0560 & 0.0952 & 0.0574 & 0.0974 \\
			\hline
			CsiLocal w/o Pipeline & 0.0583 & 0.1160 & 0.0597 & 0.1184 \\
			\hline
			Environments & \multicolumn{4}{c|}{Outdoor} \\
			\hline
			Encoding Dimension & \multicolumn{2}{c|}{256} & \multicolumn{2}{c|}{512} \\
			\hline
			Batch Size per UE & 400   & 800   & 400   & 800 \\
			\hline
			CsiLocal with Pipeline & 0.0561 & 0.0951 & 0.0562 & 0.0972 \\
			\hline
			CsiLocal w/o Pipeline & 0.0582 & 0.1160 & 0.0596 & 0.1184 \\
			\hline
		\end{tabular}%
		\label{tab:pipeline}%
	\end{table}%
	
	To implement the pipeline parallelism with two micro-batches, the decoder tail is divided into two sub-models by splitting at the connection between two CRBlocks. Each sub-model is allocated to an independent GPU accelerator. 
	For the ``no pipeline'' baseline, the entire decoder tail is deployed on a single GPU accelerator. By setting the encoding dimension (i.e., $c_1 = c_2$) to $256$ and $512$, we compare the wallclock durations of the ``pipeline'' and ``no pipeline'' methods using indoor CSI data and outdoor CSI data. 
	The results are presented in Table~\ref{tab:pipeline}.
	When the mini-batch size per UE  is set as $400$ and encoding dimension is set as $256$, wallclock durations for training $20,000$ iterations is approximately reduced by $4.11\%$ for indoor data and $3.74\%$ for outdoor data. 
	When the mini-batch size per UE is set as $800$ and encoding dimension is set as $256$, wallclock durations for training $20,000$ iterations is approximately reduced by $21.85\%$ for indoor data and $21.98\%$ for outdoor data. 
	We have the similar observations when setting the encoding dimension as $512$. 
	Therefore, we conclude that the wallclock reduction can become even bigger when the mMIMO system has more UEs and larger mini-batch size.
	We will use the mini-batch size as $800$ for the remaining numerical experiments. 
	
	\subsection{Impacts of Compression Ratio}
	\begin{figure}[!h]
		\centering
		\includegraphics[width=\linewidth]{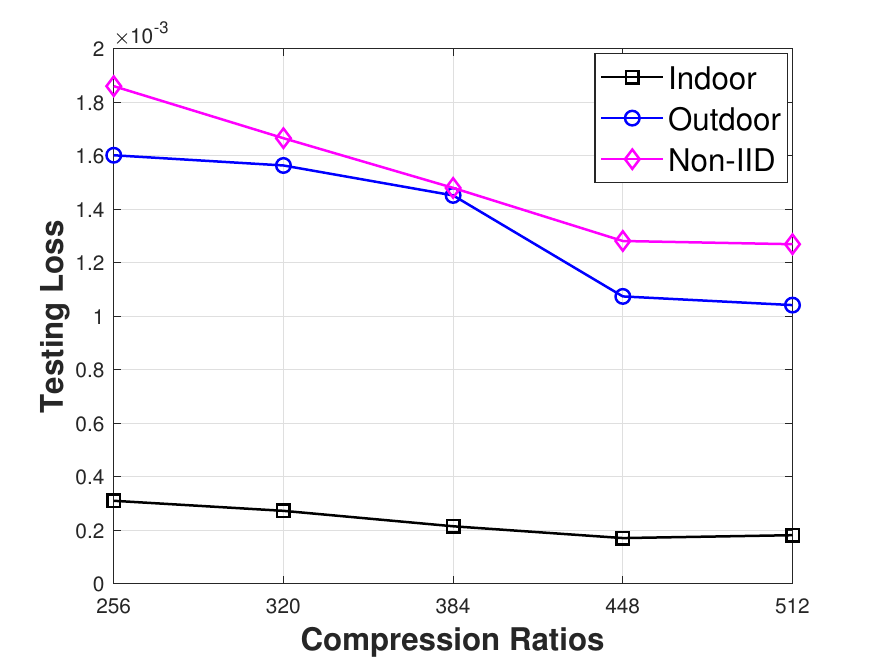}
		\caption{The impact of compression ratios on the CSI reconstruction accuracy.}
		\label{fig:compress-ratio}
	\end{figure}

	Figure \ref{fig:compress-ratio} illustrates the impact of varying compression ratios on the testing loss across three distinct data distributions: Indoor, Outdoor, and Non-IID. 
	We observe  that the proposed CSILocal algorithm effectively reduces testing loss and underscores its ability to balance the trade-off between CSI reconstruction accuracy and communication efficiency. 
	Notably, the Non-IID setting consistently exhibits the highest testing loss, which highlights the adverse effects of data heterogeneity on model performance. 
	Furthermore, the results in Fig. \ref{fig:compress-ratio} confirm that increased compression ratios generally lead to improved reconstruction accuracy with the degradation being particularly pronounced under Non-IID conditions.
	
\begin{figure*}[ht]
	\centering
	\includegraphics[width=0.41\linewidth]{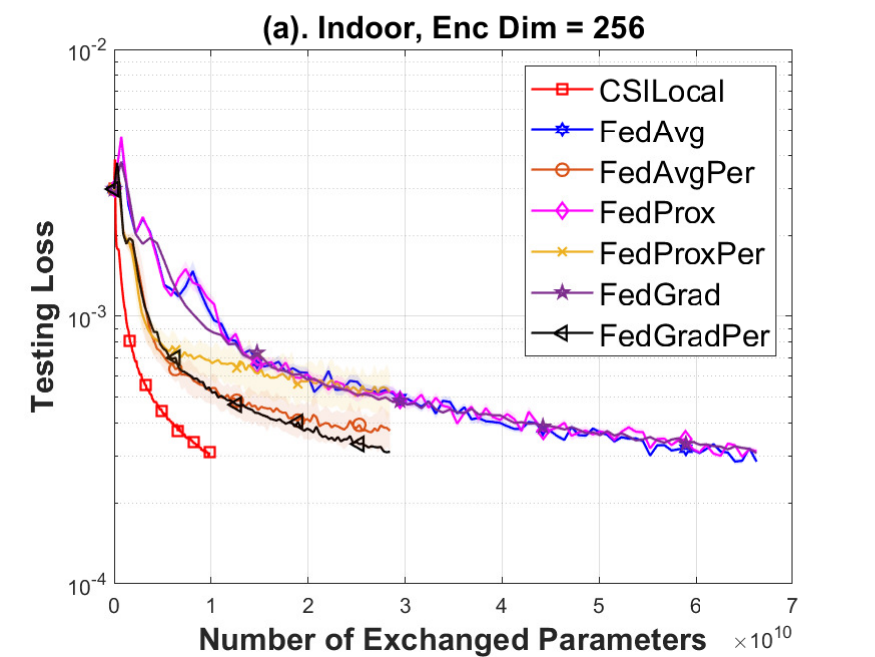}
	\includegraphics[width=0.41\linewidth]{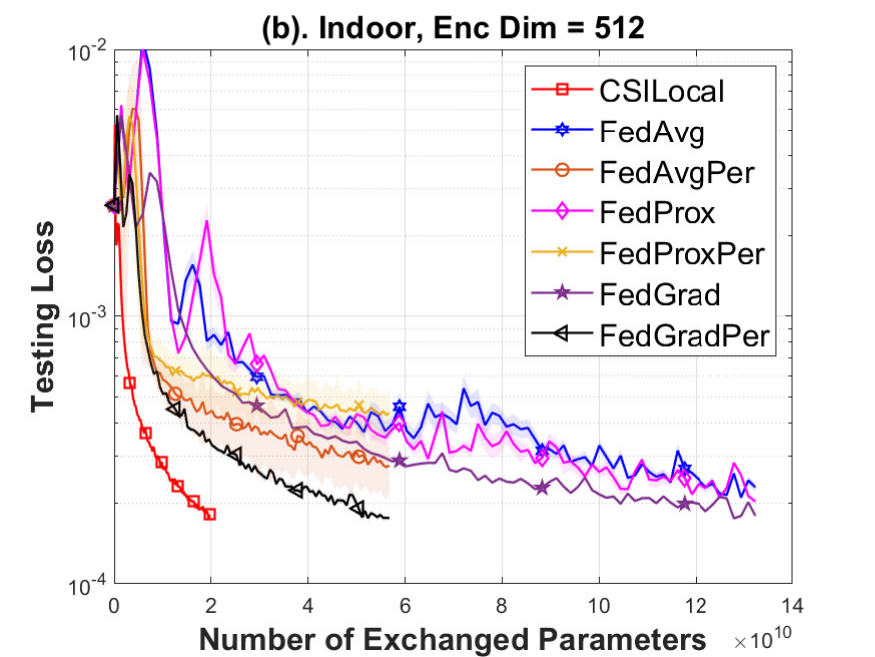}
	\includegraphics[width=0.41\linewidth]{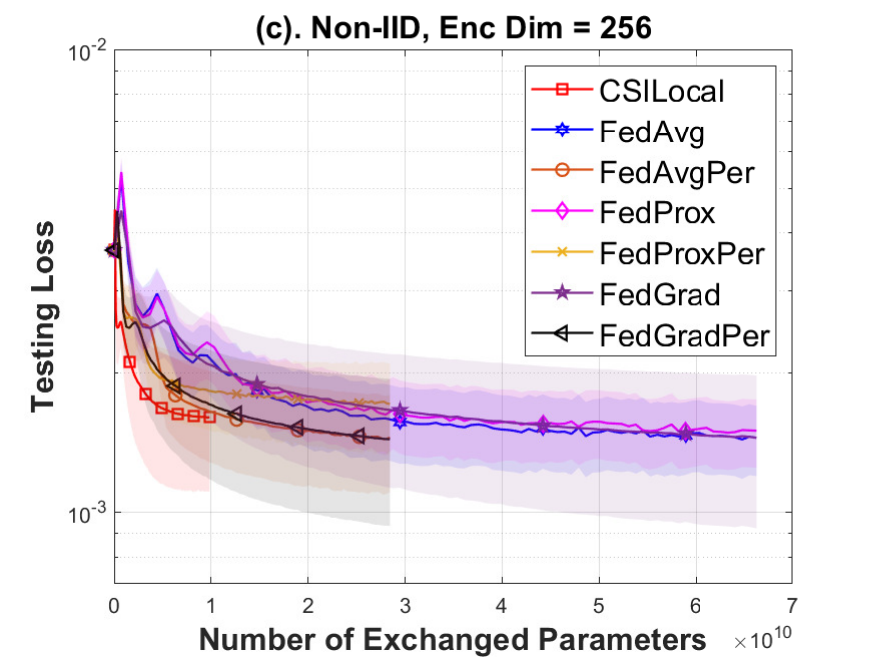}
	\includegraphics[width=0.41\linewidth]{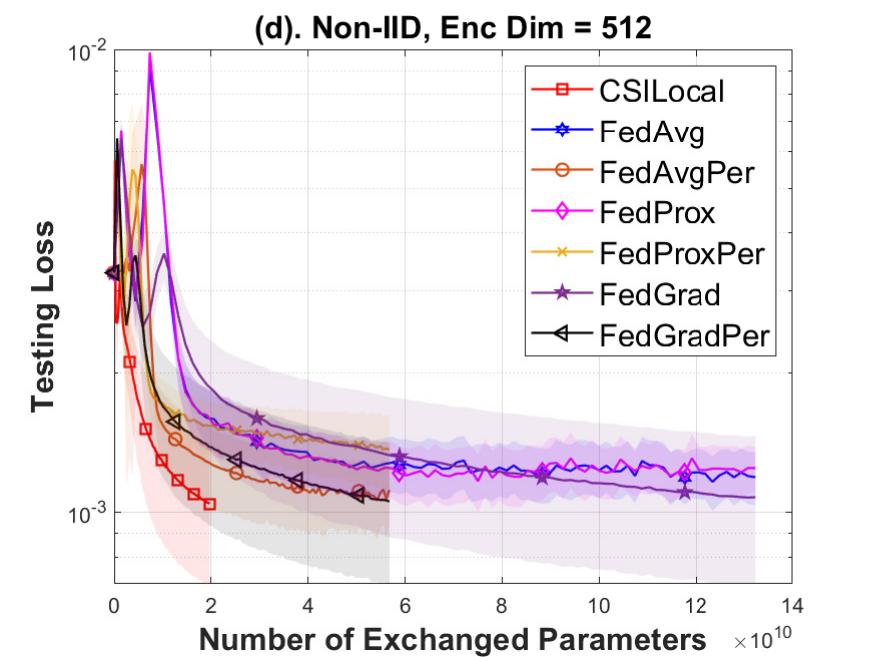}
	\includegraphics[width=0.41\linewidth]{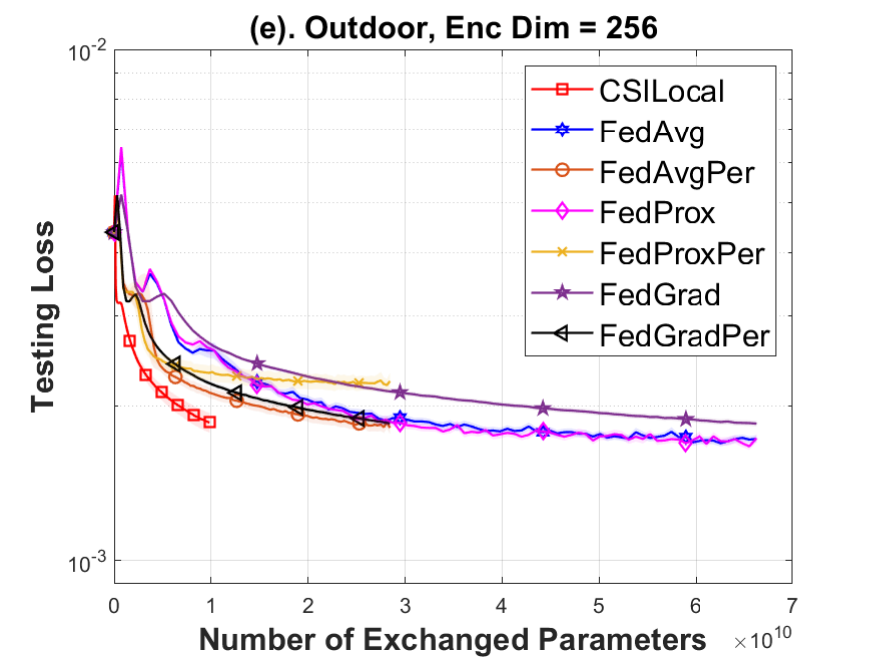}
	\includegraphics[width=0.41\linewidth]{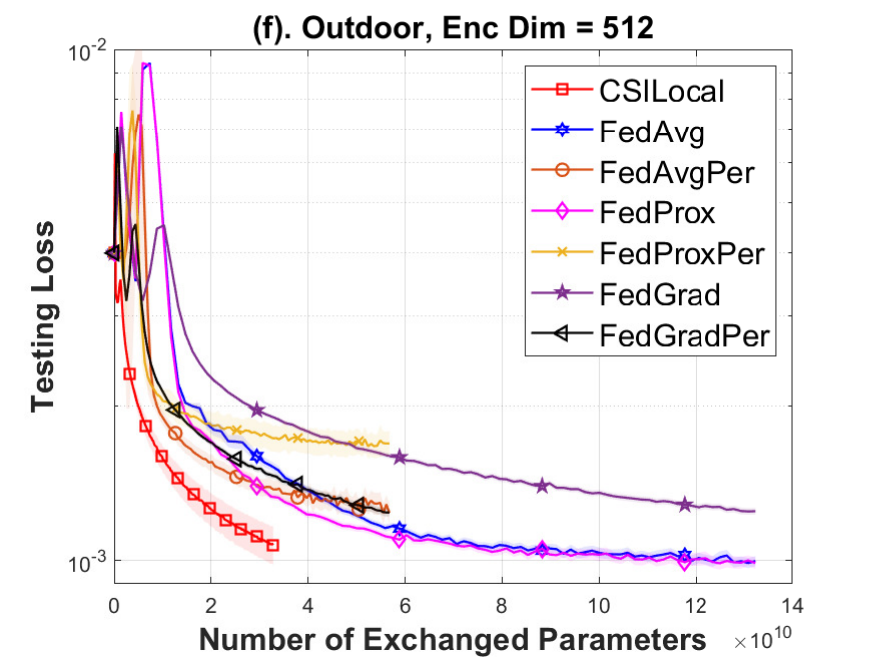}
	\caption{The convergence of testing loss over the communication overhead for the different environments and encoding dimensions.}
	\label{fig:indoor256}
	\vspace{-0.3 cm}
	\hrulefill
	\vspace{-0.5 cm}
\end{figure*}

	\subsection{Communication Efficiency}
	Figure \ref{fig:indoor256} illustrates the convergence of testing loss with respect to the number of exchanged parameters under three distinct data distributions: Indoor, Outdoor, and Non-IID. 
	We observe from Fig. \ref{fig:indoor256} that our proposed CSILocal algorithm can always achieve the lowest testing loss  while maintaining a similar testing loss over the benchmarks. 
	In the indoor scenario with $256$ encoding dimension (Fig. \ref{fig:indoor256}(a)), the proposed  CSILocal achieves relatively stable and low testing loss.
	In contrast, the outdoor scenario with $256$ encoding dimension (Fig. \ref{fig:indoor256}(e)) demonstrates slightly higher loss values and greater variability which reflects the increased channel fluctuation and reduced predictability typical of outdoor settings. 
	The most challenging scenario appears in the heterogeneous case (Fig. \ref{fig:indoor256}(c)), where data is non-IID across UEs. 
	More specifically, Fig. \ref{fig:indoor256}(c) illustrates that the heterogeneous scenario achieves a consistently higher testing loss than indoor and outdoor scenarios, which confirms the compounded effect of data heterogeneity and information loss due to compression. 
	All the three settings indicate that the proposed CSILocal algorithm can can maintain acceptable reconstruction performance under low-to-moderate compression. 
	These insights validate CSILocal's robustness in various channel conditions and its potential limitations when handling heterogeneous UE data. 
	The non-IID scenario can experience a more severe CSI reconstruction degradation over the indoor and outdoor scenarios. 
	Besides, the non-IID data scenario can induce higher fluctuation of testing loss across different UEs over the indoor and outdoor scenarios. 
	However, as shown in Fig. \ref{fig:indoor256}(b), Fig. \ref{fig:indoor256}(d), and Fig. \ref{fig:indoor256}(f), the CSI reconstruction degradation and the fluctuation of testing loss can be compensated by increasing encoding dimensions. 
	Although FedAvg and FedProx outperform our CSILocal in Fig. \ref{fig:indoor256}(f), they require significantly larger volume of exchanged parameters. Specifically, CSILocal necessitates only 19.66 billion exchanged parameters, whereas FedAvg and FedProx respectively require 47.06 billion and 38.24 billion parameters to achieve the same testing loss (e.g., $1.3 \times 10^{-3}$).

	\section{Concluding Remarks}
	A novel communication-efficient model splitting algorithm (i.e., CSILocal) was presented to reduce the communication overhead between the UEs and the BTS in the mMIMO systems. 
	The CSILocal algorithm allows each UE to exchange only local smashed data with the BTS, and thereby significantly minimizes the communication overhead while preserving local data privacy.
	To further optimize system performance, the pipeline parallelism was integrated into the decoder tail at the BTS. 
	The pipeline parallelism allows efficient processing of the accumulated smashed data and reduces the overall wallclock duration required for model training. 
	By distributing the computational tasks across multiple processing stages, pipeline parallelism can enhance the scalability of the system.
	Empirical evaluations had demonstrated that the proposed approach achieves substantial reductions in wallclock duration with the observed reduction ratio increasing proportionally to both the number of UEs involved in the system and the mini-batch size allocated per UE. 
	These results highlighted the scalability and effectiveness of the proposed model-splitting algorithm in large-scale mMIMO deployments.

	\balance
	\bibliographystyle{IEEEtran}
	\bibliography{new_RL}

\end{document}

%% file: macros.tex
\usepackage{bm}
\usepackage{nicefrac} 
\usepackage{graphicx}
\usepackage{booktabs}
\usepackage{amsfonts}
\usepackage{amsmath}
\usepackage{amssymb}
\usepackage{pifont}
\usepackage{nicefrac} 
\usepackage{algorithm}
\usepackage{algorithmicx}
\usepackage{algpseudocode}
\usepackage{dsfont}
\usepackage{authblk}
\usepackage{bbm}
\usepackage{graphicx}
\usepackage{epstopdf}
\usepackage{subfigure}
\usepackage{stfloats}
\usepackage{eqnarray}
\usepackage{makecell}
\usepackage{multirow}
\usepackage{enumerate}
\usepackage{booktabs}
\usepackage{cite}
\usepackage{balance}
\usepackage{color}
\usepackage{bm}
\usepackage{caption}
\usepackage{mathtools}
\usepackage{url}

\newcommand{\cc}{\mathbb{C}}

\newcommand{\norm}[1]{\|#1\|}

\newcommand{\adam}[1]{{\mbox{Adam}(#1)}}

\newtheorem{remark}{\textbf{Remark}}